# Switchable anomalous Hall effects in polar-stacked 2D antiferromagnet MnBi$_2$Te$_4$

Tengfei Cao[*], Ding-Fu Shao[*,†], Kai Huang, Gautam Gurung[‡], and Evgeny Y. Tsymbal[*]

*Department of Physics and Astronomy & Nebraska Center for Materials and Nanoscience,*
*University of Nebraska, Lincoln, Nebraska 68588-0299, USA*

Van der Waals (vdW) assembly allows controlling symmetry of two-dimensional (2D) materials that determines their physical properties. Especially interesting is the recently demonstrated breaking inversion symmetry by polar layer stacking to realize novel electronic, magnetic, and transport properties of 2D vdW materials switchable by induced electric polarization. Here, based on symmetry analyses and density-functional calculations, we explore the emergence of the anomalous Hall effect (AHE) in antiferromagnetic MnBi$_2$Te$_4$ films assembled by polar layer stacking. We demonstrate that breaking $\hat{P}\hat{T}$ symmetry in an MnBi$_2$Te$_4$ bilayer makes this 2D material magnetoelectric and produces a spontaneous AHE switchable by electric polarization. We find that reversable polarization at one of the interfaces in a three-layer MnBi$_2$Te$_4$ film drives a metal-insulator transition, as well as switching between an AHE and quantum AHE (QAHE). Finally, we predict that engineering an interlayer polarization in a three-layer MnBi$_2$Te$_4$ film allows converting MnBi$_2$Te$_4$ from a trivial insulator to a Chern insulator. Overall, our work emphasizes the emergence of quantum-transport phenomena in 2D vdW antiferromagnets by polar layer stacking, which do not exist in this material in the bulk or bulk-like thin-film forms.

Symmetry breaking plays an important role in generating quantum states in condensed matter, and thus its control allows manipulating these states and the associated transport properties, promising new functionalities[1]. For example, time-reversal ($\hat{T}$) symmetry breaking directly affects the spin density distribution in material systems, producing spontaneous magnetic orders and novel spin-dependent transport properties useful for electronic applications[2,3,4,5,6,7,8]. The control of these states usually requires an external magnetic field as the stimulus directly modulating the $\hat{T}$ symmetry breaking by reorienting the magnetic moments. However, the use of an external magnetic field in electronic devices is not efficient. On one hand, it requires substantial electric currents and thus costs large energy dissipation. On the other hand, an external magnetic field can hardly control the magnetic order in antiferromagnets – materials that have recently attracted increasing interest due to their potential application for next generation spintronics[9,10,11,12,13,14]. It would be desirable to have a method to directly control the spin-dependent properties of a material without changing its magnetic ground state, e.g., an applied electric field, although such a "nonmagnetic" method does not directly influence $\hat{T}$ symmetry [15,16,17].

In recent years, two-dimensional (2D) van der Waals (vdW) materials have become an extensively studied model system for exploring the new physics and potential applications at the nanoscale[18,19,20,21,22,23,24]. A particular interest was stimulated by the recent discoveries of unexpected phenomena driven by interlayer stacking [25,26,27,28,29]. Despite weak interlayer coupling, minor interlayer sliding[26,27,28] or small-angle interlayer twisting[25,30,31] can generate electronic and transport properties which do not exist in the bulk-like phases. This is largely due to the modification of the interlayer stacking pattern breaking crystal symmetries. Specifically, a combination of interlayer sliding and 180° twisting results in a non-centrosymmetric stacking pattern accompanied by the emergence of an out-of-plane electric polarization reversable by an external electric field [26,27,28, 32]. This approach allows the design of 2D ferroelectrics out of parent nonpolar compounds to uncover new functionalities not existent in the bulk phase[33].

Such vdW polar stacking can be employed to engineer recently discovered 2D antiferromagnets [34]. Breaking space inversion symmetry in these materials could potentially induce novel spin-dependent properties controlled by the intrinsic ferroelectric polarization. Especially interesting are 2D materials derived from antiferromagnet MnBi$_2$Te$_4$ which exhibits non-trivial transport properties in the thin-film form[4,5,35,36,37,38,39,40]. Although $\hat{T}$ symmetry is globally broken by magnetism of these antiferromagnetic materials, the magnetism is "hidden" by $\hat{T}\hat{O}$ symmetry that combines $\hat{T}$ and crystal symmetry $\hat{O}$ such as space inversion ($\hat{P}$) or mirror reflection ($\hat{M}$). Since the polar layer stacking affects crystal symmetry $\hat{O}$, it allows an indirect control of the $\hat{T}$ symmetry breaking in 2D magnets and thus spin-dependent properties, such as the anomalous Hall effect (AHE).

The intrinsic AHE is governed by the Berry curvature $\boldsymbol{\Omega}$, a property arising from the spin-dependent band structure of magnetic systems [41]. The anomalous Hall conductance (AHC) is quantized in 2D magnetic topological insulators due to formation of the topologically protected edge channels resulting in a dissipationless quantum AHE (QAHE) [2,4,35]. Ferromagnetic systems naturally support the AHE due to the intrinsic $\hat{T}$ symmetry breaking. Antiferromagnets also break $\hat{T}$ symmetry, but usually preserve $\hat{T}\hat{O}$ symmetry that enforces the Berry curvature to be antisymmetric in *k*-space, i.e. $\hat{T}\hat{O}\boldsymbol{\Omega}(\boldsymbol{k}) = -\boldsymbol{\Omega}(\hat{T}\hat{O}\boldsymbol{k})$, prohibiting the AHE. The AHE can emerge in magnetically uncompensated antiferromagnets, where the combined $\hat{T}\hat{O}$ symmetry is broken either by noncollinear magnetic configurations or non-magnetic atoms at low symmetry positions [6,41,42,43,44,45]. In the former case, the AHE is dependent on the orientation of the Néel vector, which can be changed by an applied magnetic field. In the latter case, the atomic structure of the non-magnetic sublattice can influence the AHE; however, there are no straightforward means to control this structure by external stimulus.

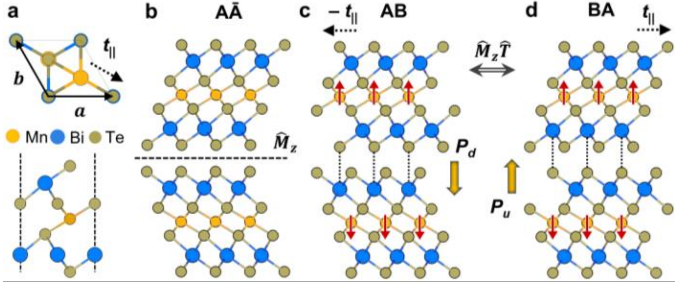

**Fig. 1** (**a**) Atomic structure of monolayer MnBi$_2$Te$_4$ (top and side views). (**b**) Unstable AĀ stacking of a MnBi$_2$Te$_4$ bilayer where the top monolayer is vdW assembled as a mirror reflection ($\widehat{M}_z$) of the bottom monolayer. (**c, d**) A polar-stacked MnBi$_2$Te$_4$ bilayer with antiparallel magnetic moments and electric polarization pointing down $P_d$ (c) and up $P_u$ (d) due to the in-plane translation, $-t_\parallel$ (c) or $t_\parallel$ (d), of the top monolayer from the AĀ stacking in (b). The bilayer structures with opposite polarization are related by the $\widehat{M}_z\widehat{T}$ symmetry operation or equivalently by in-plane translation $2t_\parallel$.

In this letter, based on symmetry analysis and first-principles density functional calculations, we demonstrate that the AHE can be effectively induced in 2D vdW antiferromagnets by polar layer stacking. Using recently discovered antiferromagnetic topological insulator MnBi$_2$Te$_4$, as a representative material, we show that polar layer stacking breaks $\widehat{P}\widehat{T}$ symmetry and induces a magnetoelectric effect, a normal AHE, or QAHE, depending on the number of MnBi$_2$Te$_4$ layers and their stacking. These effects can be controlled by the switchable electric polarization of the polar stacked MnBi$_2$Te$_4$, thus providing a route to manipulate the spin-dependent and quantum properties without an external magnetic field.

**Results and discussion**

Monolayer MnBi$_2$Te$_4$ represents a septuple layer, which can be viewed as a MnTe bilayer intercalated into the center of a Bi$_2$Te$_3$ quintuple layer (Fig. 1(a)). Bulk MnBi$_2$Te$_4$ has $\widehat{P}\widehat{T}$ symmetry and an A-type antiferromagnetic order with colinear out-of-plane magnetic moments. A MnBi$_2$Te$_4$ film in bulk-like stacking inherits the bulk A-type antiferromagnetic order and exhibits quantum phenomena dependent on the number of layers [2,4,36]. An MnBi$_2$Te$_4$ film with an even number of layers, e.g., a bilayer, has intrinsic $\widehat{P}\widehat{T}$ symmetry, which not only prevents the net magnetic moment, but also impedes both the AHE and QAHE. This can be seen from the expression for the Berry curvature:

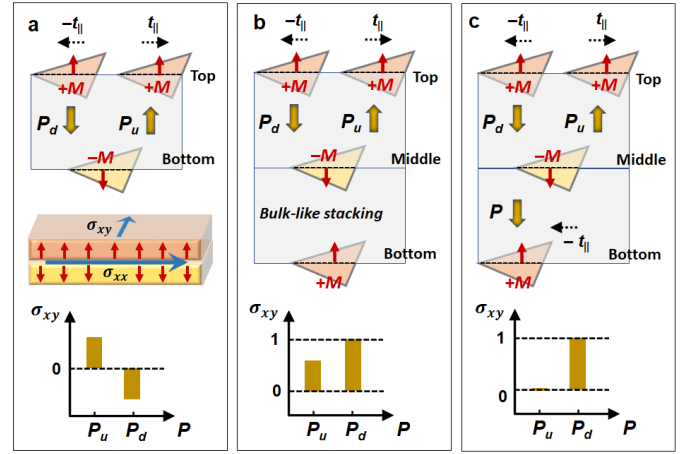

**Fig. 2** (**a**) Top: schematic of a polar-stacked MnBi$_2$Te$_4$ bilayer with polarization pointing up (**P$_u$**) or down (**P$_d$**). Middle and bottom: an AHE in a polar-stacked MnBi$_2$Te$_4$ bilayer switchable by ferroelectric polarization. (**b**) Top: schematic of a MnBi$_2$Te$_4$ trilayer with a single (top) polar interface. Bottom: switching between AHE and QAHE with polarization reversal. (**c**) Top: schematic of a MnBi$_2$Te$_4$ trilayer with two polar interfaces. Bottom: transition between trivial and Chern insulators driven by polarization switching.

$$\Omega(\boldsymbol{k}) = -2\mathrm{Im}\sum_{m\neq n}\frac{\langle n|\frac{\partial \widehat{H}}{\partial k_x}|m\rangle\langle m|\frac{\partial \widehat{H}}{\partial k_y}|n\rangle}{(E_{n\boldsymbol{k}}-E_{m\boldsymbol{k}})^2}, \quad (1)$$

where $\widehat{H}$ is the Hamiltonian of the system and $E_{n\boldsymbol{k}}$ is the energy of the $n$-th band at wave vector $\boldsymbol{k}$. The AHC determined by the integration of $\Omega(\boldsymbol{k})$ in the Brillouin zone (BZ) as follows:

$$\sigma_{xy} = -\frac{e^2}{h}\int_{BZ}\frac{d^2\boldsymbol{k}}{2\pi}\Omega(\boldsymbol{k}). \quad (2)$$

According to Eq. (1), the Berry curvature is odd under the $\widehat{P}\widehat{T}$ symmetry operation, i.e., $\widehat{P}\widehat{T}\Omega(\boldsymbol{k}) = -\Omega(\boldsymbol{k})$, and thus is zero in any $\widehat{P}\widehat{T}$-symmetric system enforcing $\sigma_{xy}$ to vanish. On the contrary, due to the A-type antiferromagnetic order, an MnBi$_2$Te$_4$ film with an odd number of layers does not have the restriction of $\widehat{P}\widehat{T}$ symmetry, due to an uncompensated net magnetic moment. As a result, an intrinsic QAHE can occur in such a film where the number of quantum states is determined by the number of MnBi$_2$Te$_4$ layers, as was predicted theoretically [4] and demonstrated experimentally [2,36,38]. Here, we show that a polar-stacked MnBi$_2$Te$_4$ provides even a broader spectrum of quantum states and these states can be controlled by electric polarization.

Using the vdW assembly approach demonstrated in refs. 26 and 27, a MnBi$_2$Te$_4$ bilayer can be engineered to be polar. This requires top monolayer Ā to be deposited as mirror reflection ($\widehat{M}_z$) of the bottom monolayer. The resulting non-centrosymmetric AĀ stacking (Fig. 1(b)) is unstable due to the proximity of the Te atoms across the interface. It spontaneously relaxes to the lower-energy AB or BA stacking with antiparallel magnetic moments (Figs. 1(c,d)). The latter is distinguished by the lateral translation, $-t_\parallel$ or $t_\parallel$, along the [1$\bar{1}$0] direction of the top MnBi$_2$Te$_4$ monolayer with respect to the bottom ($t_\parallel = \frac{1}{3}(a - b)$ in Fig. 1(a)), placing the interfacial Bi atom atop the Te atom. The resulting two structures, AB and BA, have opposite polarizations pointing down **P$_d$** or up **P$_u$** depending on the lateral translation, $-t_\parallel$ or $t_\parallel$, respectively (Figs. 1(c,d)). The polarization can be switched by an applied electric field through interlayer sliding.

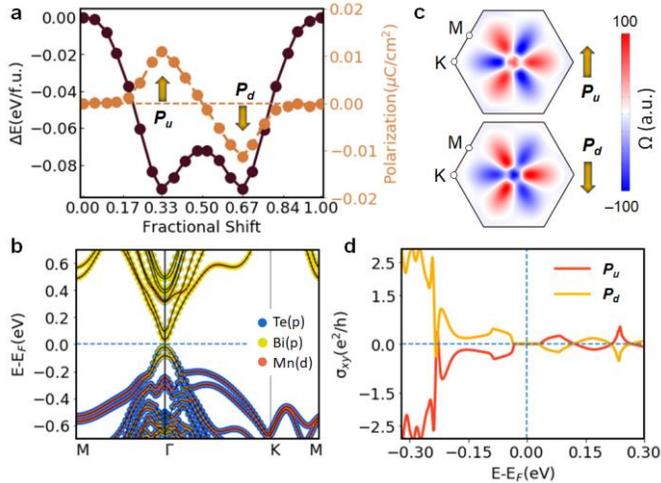

**Fig. 3** (**a**) Total energy and out-of-plane electric polarization of a polar MnBi$_2$Te$_4$ bilayer when sliding the top monolayer with respect to the bottom along the [1$\bar{1}$0] direction. The two minima in energy profile correspond to polarization pointing up (**P$_u$**) and down (**P$_d$**). (**b**) Electronic band structure of a polar MnBi$_2$Te$_4$ bilayer along the high-symmetry directions in the 2D Brillouin zone. (**c**) Berry curvature of a polar MnBi$_2$Te$_4$ bilayer in the 2D Brillouin zone calculated at $E - E_F = 0.07$ eV for polarization pointing up (top) and down (bottom). (**d**) Anomalous Hall conductance $\sigma_{xy}$ of a polar MnBi$_2$Te$_4$ bilayer as a function of electron energy for polarization pointing up and down.

Depending on the number of monolayers, the interlayer polarization has different effects on quantum states of MnBi$_2$Te$_4$ films. In case of a *bilayer* (Fig. 2 (a)), polar staking breaks inversion symmetry and hence $\hat{P}\hat{T}$ symmetry. This lifts the $\hat{P}\hat{T}\Omega(\mathbf{k}) = -\Omega(\mathbf{k})$ constraint on the Berry curvature of the bulk phase making AHC of a polar MnBi$_2$Te$_4$ bilayer nonzero. The two polarization states are related by the $\hat{M}_z\hat{T}$ symmetry transformation (Fig. S2(a)), which changes sign of the Berry curvature, $\hat{M}_z\hat{T}\Omega(\mathbf{k}) = -\Omega(-\mathbf{k})$, and hence sign of the AHC $\sigma_{xy}$ in Eq. (2). Therefore, the switchable polarization of the polar MnBi$_2$Te$_4$ bilayer can serve as a control parameter for the AHE.

For a polar-stacked MnBi$_2$Te$_4$ *trilayer*, the net magnetic moment is non-zero and hence the AHE is always finite due to broken $\hat{P}\hat{T}$ symmetry. In case of a trilayer with a polar MnBi$_2$Te$_4$ bilayer deposited on an MnBi$_2$Te$_4$ monolayer with bulk-like interface (Fig. 2(b)), the interlayer polarization controls band bending across the layer stack and can lead to the AHE or QAHE depending on the polarization orientation. In case of a trilayer with two polar interfaces (Fig 2(c)), the polarization of the top interface can be switched to be parallel or antiparallel to the polarization of the bottom interface resulting in the topological phase transition from a trivial band insulator to a non-trivial Chern insulator.

To explicitly demonstrate the anticipated properties, we apply density functional calculations, as described in Supplementary Material. Fig. 3 (a) shows the calculated energy profile and corresponding polarization for a polar MnBi$_2$Te$_4$ bilayer when sliding the top layer with respect to the bottom along the [1$\bar{1}$0] direction (see Fig. S1 for the sliding energy profile along the [100] direction). Zero fractional shift matches to the A$\bar{\text{A}}$ stacking which has the highest energy and thus unstable. The energy drops down with the shift and exhibits two local minima separated by an energy barrier of 0.09 eV/f.u. These minima occur at a fractional shift of a third and two thirds the in-plane unit cell vector in the [1$\bar{1}$0] direction (equivalent to the lateral translation, $t_\parallel$ and $-t_\parallel$, respectively). The two local energy minima correspond to the polar structures with polarization pointing up and down (Fig. 3(a)). The interlayer polarization is mainly contributed by the charge transfer and associated displacements of the interfacial cation Bi and anion Te atoms, creating a dipole pointing from Bi to Te. The calculated out-of-plane polarization magnitude is 0.01 μC/cm$^2$. This polarization induces the electrostatic potential energy drop across the MnBi$_2$Te$_2$ bilayer of 0.05 eV as shown in Fig. S2 (b)). Thus, the interlayer sliding transforms the polar structure with polarization pointing up to the structure with polarization pointing down or vice versa. Such sliding and the associated polarization reversal can be induced by an applied out-of-plane electric field of less than 1 V/nm, which is feasible in experimental conditions.

The $\hat{P}\hat{T}$ symmetry broken by polar-layer stacking breaks Kramers' degeneracy of the energy bands for the bilayer. In the absence of spin-orbit coupling (SOC), the bands are exchange split into spin-up ($\sigma = \uparrow$) and spin-down ($\sigma = \downarrow$) states (Fig. S3). The spin character of the bands $E^\sigma(\mathbf{k})$ is reversed with ferroelectric polarization, as follows from $\hat{M}_z\hat{T}E^\uparrow(\mathbf{k}) = E^\downarrow(-\mathbf{k})$. Including SOC, changes the band structure of a polar MnBi$_2$Te$_4$ bilayer. As seen from Fig. 3(b), similar to bulk MnBi$_2$Te$_4$, the polar bilayer represents a direct band gap semiconductor with the conduction band minimum (CBM) and the valence band maximum (VBM) located at the Γ point. The states around the Fermi level are mainly composed of the *p* states of Bi and Te with the *d* states of Mn being relatively far away from the Fermi energy. Due to the presence of ferroelectric polarization, the band gap of the polar bilayer is reduced to about 0.06 eV from 0.08 eV obtained for a non-polar bilayer and 0.16 eV known for bulk MnBi$_2$Te$_4$.

We note that a polar-stacked MnBi$_2$Te$_4$ bilayer represents a magnetoelectric multiferroic. It has both a spontaneous electric polarization and an antiferromagnetic order. The magnetic moments of Mn atoms of about 4.47 μ$_B$ are aligned antiparallel on top and bottom MnBi$_2$Te$_4$ monolayers. Importantly, due to broken $\hat{P}\hat{T}$ symmetry, the antiferromagnetism of the polar bilayer is not fully compensated. While the uncompensated net magnetic moment is small, about 0.002 μ$_B$ per formular unit, it is non-vanishing and reversible by ferroelectric polarization, as follows from $\hat{M}_z\hat{T}\boldsymbol{M} = -\boldsymbol{M}$, where $\boldsymbol{M}$ is the net magnetization. Potentially this functionality of the polar-stacked vdW antiferromagnets may be useful for applications.

The broken $\hat{P}\hat{T}$ symmetry causes the Berry curvature of a polar-stacked MnBi$_2$Te$_4$ bilayer to be non-vanishing. Fig. 3(c) shows the calculated Berry curvature $\Omega(\boldsymbol{k})$ in the 2D Brillouin zone for up and down polarization states at energy $E - E_F = 0.07$ eV above the band gap. It is seen that $\Omega(\boldsymbol{k})$ exhibits a flower-like pattern with alternating red and blue color of the petals (i.e. alternating sign of $\Omega(\boldsymbol{k})$) reflecting the 3-fold rotational symmetry of the trilayer around the z axis. This symmetry combined with $\hat{M}_z\hat{T}\Omega(\boldsymbol{k}) = -\Omega(-\boldsymbol{k})$, makes the color of the petals (reflecting the sign of $\Omega(\boldsymbol{k})$) independent of the polarization orientation. On the contrary, a persistent color around the Γ point is reversed from red to blue when polarization is switched from pointing up to down.

The non-vanishing Berry curvature implies finite AHC. Fig. 3(d) shows the calculated AHC as a function of electron energy. It is evident that the AHC is zero within the band gap of an MnBi$_2$Te$_4$ bilayer, indicating that the bilayer represents a trivial insulator (semiconductor). However, with electron or hole doping, AHC becomes non-zero and, as expected, has opposite sign for ferroelectric polarization pointing up and down. Thus, we observe the new functionality of a polar-stacked MnBi$_2$Te$_4$ bilayer which does not exist in bulk-like MnBi$_2$Te$_4$.

The presence of a finite AHC in a polar-stacked MnBi$_2$Te$_4$ bilayer can be understood in terms of a layer-dependent Hall effect[39]. While a bulk-like MnBi$_2$Te$_4$ bilayer represents an axion insulator where the top and bottom MnBi$_2$Te$_4$ monolayers spontaneously deflect electrons in opposite directions resulting in zero net AHE, the presence of spontaneous polarization in a polar MnBi$_2$Te$_4$ bilayer breaks the AHE-compensating symmetry between the top and bottom monolayers producing a finite AHC. This phenomenon is analogous to the effect of electric field on a bulk-like MnBi$_2$Te$_4$ bilayer[39]. However, while in the case of a bulk-like MnBi$_2$Te$_4$ bilayer, the electric field needs to be maintained to have a non-zero AHC, the AHE in a polar MnBi$_2$Te$_4$ bilayer is spontaneous and non-volatile with the sign of AHC being linked to the direction of electric polarization.

The polarization controlled AHC emerges not only in a MnBi$_2$Te$_4$ bilayer that is composed of two polar-stacked MnBi$_2$Te$_4$ monolayers, but also in MnBi$_2$Te$_4$ films assembled by polar stacking of two MnBi$_2$Te$_4$ layers formed of two or more monolayers. Fig. S3 shows the results of calculations of the band structure and AHC for polar stacked four- (Fig. S3(a)) and six- (Fig. S3(d)) monolayer MnBi$_2$Te$_4$ films. It is seen that compared to the MnBi$_2$Te$_4$ bilayer, the band gap is reduced in the four-monolayer MnBi$_2$Te$_4$ (Fig. S3(b)) and almost vanishes in the six-monolayer MnBi$_2$Te$_4$ (Fig. S3(e)). The AHC as a function of energy (Figs. S3(c,f)) exhibits trends similar to those of the bilayer. In particular, the AHC is reversed by switching the interlayer polarization. Thus, in experimental conditions, vdW assembly can be used to engineer polar MnBi$_2$Te$_4$ films with an even number of monolayers to realize an AHE switchable by the intrinsic ferroelectric polarization.

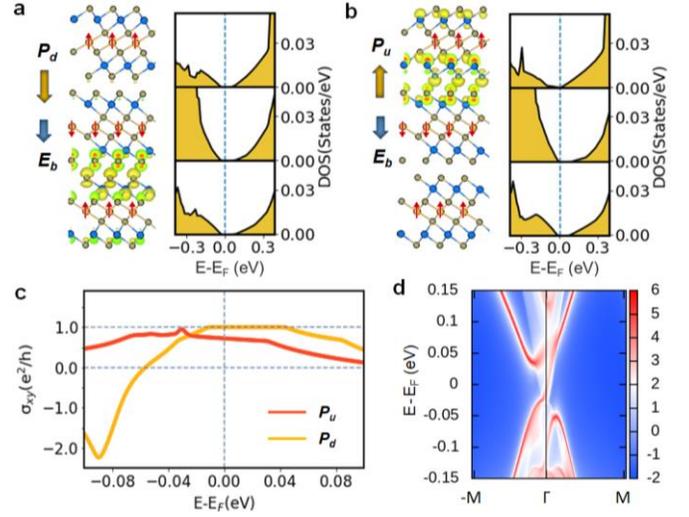

**Fig. 4** (**a,b**) Three-monolayer MnBi$_2$Te$_4$ film with a polar top interface with polarization pointing down ($\boldsymbol{P_d}$) (a) or up ($\boldsymbol{P_u}$) (b) and non-polar bulk-like bottom interface (left panels), and corresponding layer-resolved density of states (DOS) as a function of energy(right panels). In the left panels, blue arrows indicate a built-in electric field due to the band offset between top and bottom interfaces in the trilayer. The real-space distributions of the charge density at the conduction band minimum are shown by yellow colored isosurfaces of 10 % of their maxima. (**c**) Anomalous Hall conductivity as a function of electron energy for $\boldsymbol{P_u}$ and $\boldsymbol{P_d}$. (**d**) Surface band structure of the three-monolayer MnBi$_2$Te$_4$ film for $\boldsymbol{P_u}$ indicating the appearance of the topologically protected edge state at the Fermi energy.

Next, we consider effects of polar stacking in three-monolayer MnBi$_2$Te$_4$ films. Due A-type antiferromagnetism, such films have a large uncompensated magnetic moment of about 4.47 μ$_B$ per supercell. The presence of this magnetic moment breaks $\hat{P}\hat{T}$ symmetry of the parent bulk phase producing a nonzero Berry curvature and the associated AHE. Following the schematic Fig. 2(b), we assume that a three-monolayer MnBi$_2$Te$_4$ slab is composed of a polar-stacked bilayer deposited on a MnBi$_2$Te$_4$ monolayer with the bulk-like stacked bottom interface (Figs. 4(a,b) left panels)). We find a small band offset (~ 0.05 eV) between the polar and non-polar stacked interfaces resulting in a built-in electric field $\boldsymbol{E_b}$ across the trilayer pointing down (indicated by blue arrows in Figs.

4(a,b)). The presence of this bias field independent of polarization orientation makes the transport behavior of the polar MnBi$_2$Te$_4$ trilayer different from that of a polar bilayer.

When polarization is pointing down ($P_d$), the associated depolarizing field is directed opposite to the bias field and effectively compensates it, thus maintaining the band gap across the whole polar trilayer (≈ 0.07 eV), as seen from the layer-resolved density of states (DOS) in Fig. 4(a). Also, as evident from the real space distribution of the charge density at the CBM in this figure, the CBM lies away from the polar interface. It is expected in this case that the transport behavior of the polar trilayer in the $P_d$ state to be reminiscent to that of the bulk-like non-polar trilayer. The latter is known to represent a Chern insulator exhibiting an QAHE[5]. Likewise, we find that the polar MnBi$_2$Te$_4$ trilayer in the $P_d$ state signifies a Chern insulator with the Chern number equal to 1 and exhibits an QAHE with AHC $\sigma_{xy} = e^2/h$ in the energy gap region (yellow curve in Fig. 4(c)). The quantized AHC is associated with the non-trivial topological character of the system which is reflected in the topologically protected edge state shown in Fig. 4 (d).

On the contrary, when polarization is pointing up ($P_u$), the associated depolarizing field is aligned in the direction of the bias field which effectively doubles its effect producing strong band bending across the trilayer, as seen from the layer-resolved DOS in Fig. 4(b). As a result, the band gap is closed at the polar interface of the trilayer, producing a non-zero charge density at the Fermi energy. The metallic character of the polar MnBi$_2$Te$_4$ trilayer in the $P_d$ state makes the Chen number equal to 0, eliminates a topologically protected surface state, and changes the AHC to a smaller value at the Fermi energy (red curve in Fig. 4(c)). Thus, the switchable electric polarization of the polar MnBi$_2$Te$_4$ trilayer can serve as a knob to control the transition between the non-trivial (associated with QAHE) and trivial (associated with AHE) electronic states in the MnBi$_2$Te$_4$ films.

A different type of stacking in a MnBi$_2$Te$_4$ trilayer occurs when a polar MnBi$_2$Te$_4$ bilayer is placed on a monolayer in such a way that the bottom interface is also polar and has polarization pointing down (Fig. 2(c)). In this case the trilayer has two polar interfaces and the polarization of the top interface can be switched up ($P_u$) to be antiparallel (AP) or down ($P_d$) to be parallel (P) to the polarization $P$ of the bottom interface (Figs. 5(a) and 5(d), respectively). Depending on the relative polarization orientation (AP or P), we find different topology of the trilayer. While in both cases, we observe the presence of a band gap in the band structure of AP- (Fig. 5(b)) and P- (Fig. 5(e)) aligned trilayers, the nature of this gap is different. By comparing the band structures in Fig. 5(b) and Fig. 5(e), we see for the former, that the bands exhibit conventional dispersions typical for a normal semiconductor with the CBM and VBM located at the Γ point. On the contrary, for the latter, the valence bands reveal a "mexican-hat" shape placing the VBM away from the Γ point and indicating a possible band inversion and thus a non-trivial character of the band gap.

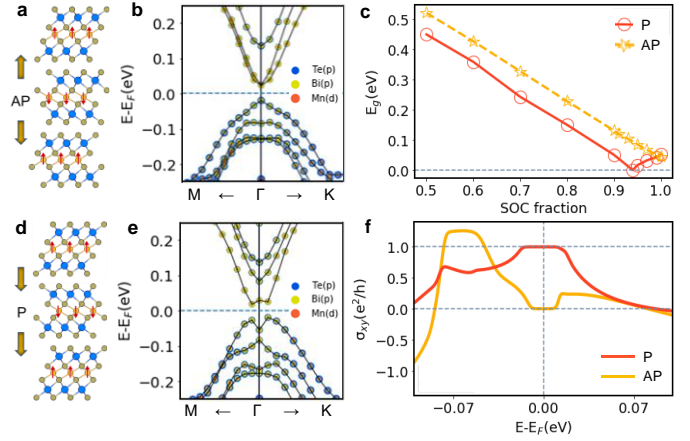

**Fig. 5:** (**a,b,d,e**) Atomic structure (a,d) and energy bands near the Γ point (b,e) of a polar MnBi$_2$Te$_4$ trilayer with antiparallel (AP) (a,b) and parallel (P) (d,e) interface polarization. (**c**) Variation of the band gap for parallel (P) and antiparallel (AP) aligned MnBi$_2$Te$_4$ trilayer as a function of SOC strength given in units of a fraction of the actual SOC. (**f**) Anomalous Hall conductivity of the P and AP-aligned MnBi$_2$Te$_4$ trilayer as a function of electron energy.

To elucidate these different behaviors, we explore the evolution of the band gap for the AP- and P- aligned MnBi$_2$Te$_4$ trilayers as a function of SOC. Fig. 5(c) shows variation of the band gap for AP- and P-aligned aligned MnBi$_2$Te$_4$ trilayers as a function of SOC strength given in units of a fraction of the actual SOC (see Figs. S5 and S6 for the associated bands structure around the Γ point). It is seen that, with the increasing fraction of SOC, the band gap of the AP-aligned trilayer continuously reduces (yellow stars in Fig. 5(c) and Fig. S5), until it reaches the smallest value of 0.06 eV at the actual value of SOC. On the contrary, the band gap of the P-aligned trilayer first decreases with increasing SOC, so that the band gap reduces to zero at SOC = 0.93 (red circles in Fig. 5(c) and Fig. S6). A further increase of SOC reopens and then enlarges the band gap. This behavior indicates band inversion in the P-aligned MnBi$_2$Te$_4$ trilayer that is induced by SOC, reflecting the topologically non-trivial origin of the band gap. This contrasts to the AP-aligned MnBi$_2$Te$_4$ trilayer where there is no band inversion with increasing SOC, thus signaling the trivial character of the system. These conclusions are confirmed by the direct calculation of the Chern number. While the Chern number is 0 for the AP-aligned trilayer, it is 1 for the P-aligned trilayer.

The non-trivial character of the AP-aligned MnBi$_2$Te$_4$ trilayer is reflected in the quantized AHC exhibiting a plateau of $\sigma_{xy} = e^2/h$ within the energy gap (red curve in Fig. 5(f)), indicating that the trilayer represents a Chern insulator exhibiting an QAHE. On the contrary, $\sigma_{xy} = 0$ in the band gap region of the P-aligned MnBi$_2$Te$_4$ trilayer (yellow curve in Fig. 5(f)), indicating that the trilayer represents a trivial insulator. In both cases, for electron energies above and below the band gap the AHC is non-zero, reflecting the broken $\hat{P}\hat{T}$ symmetry of the system, and changes with reversal of ferroelectric polarization of the top interface. Therefore, a polar MnBi$_2$Te$_4$ trilayer can be switched between the trivial insulator and Chern insulator states by reversing electric polarization of the top interface resulting in the antiparallel or parallel polarization alignment.

**Conclusions**

Overall, our results demonstrate that utilizing polar stacking of a 2D vdW antiferromagnet MnBi$_2$Te$_4$ with switchable interface polarization allows achieving new functional properties that are not available in the bulk form of this material. Specifically, due to broken $\hat{P}\hat{T}$ symmetry, a polar-stacked MnBi$_2$Te$_4$ bilayer exhibits an AHE, whose sign is determined by ferroelectric polarization orientation, as well as the magnetoelectric effect with the net magnetic moment of the bilayer switchable by polarization. The reversable polarization in a single-polar-interface MnBi$_2$Te$_4$ trilayer controls the transition between metallic and insulating phases associated with the AHE and QAHE states, respectively. Switching between antiparallel and parallel interlayer polarization states in a bi-polar MnBi$_2$Te$_4$ trilayer enforces the transition between a trivial insulator to a Chern insulator carrying a topologically protected edge state and exhibiting an QAHE.

Polar layer stacking can be employed not only for MnBi$_2$Te$_4$ considered in this work, but also for other 2D vdW antiferromagnets to propel their new functional properties. Due to small vertical ion displacements, the out-of-plane electric polarization can be switched by a small applied electric field via in-plane interlayer sliding with an ultra-low barrier. These characteristics are useful for potential applications of the polar-stacked 2D vdW antiferromagnets in data storage. We hope therefore that our theoretical results will stimulate experimental efforts to verify our predictions.

**Acknowledgments.** The work was supported by the grant DE-SC0023140 funded by the U.S. Department of Energy, Office of Science. Computations were performed at the University of Nebraska Holland Computing Center.

† *Current affiliation*: Key Laboratory of Materials Physics, Institute of Solid State Physics, HFIPS, Chinese Academy of Sciences, Hefei 230031, China

‡ *Current affiliation*: Trinity College, University of Oxford, Oxford OX1 3BH, UK

* tcao6@unl.edu, dfshao@issp.ac.cn, tsymbal@unl.edu